# A Decision Support System to Choose Optimal Release Cycle Length in Incremental Software Development Environments

Avnish Chandra Suman[1], Saraswati Mishra[1], Abhinav Anand[2]

[1]Centre for Development of Telematics, New Delhi, India
[2]Intel Technologies, Bengaluru, India

## Abstract

*In the last few years it has been seen that many software vendors have started delivering projects incrementally with very short release cycles. Best examples of success of this approach has been Ubuntu Operating system that has a 6 months release cycle and popular web browsers such as Google Chrome, Opera, Mozilla Firefox. However there is very little knowledge available to the project managers to validate the chosen release cycle length. We propose a decision support system that helps to validate and estimate release cycle length in the early development phase by assuming that release cycle length is directly affected by three factors, (i) choosing right requirements for current cycle, (ii) estimating proximal time for each requirement, (iii) requirement wise feedback from last iteration based on product reception, model accuracy and failed requirements. We have altered and used the EVOLVE technique proposed by G. Ruhe to select best requirements for current cycle and map it to time domain using UCP (Use Case Points) based estimation and feedback factors. The model has been evaluated on both in-house as well as industry projects.*

## 1. Introduction

Software Release Planning has been a classical problem. Rather than making optimal release policies, vendors now lean towards getting best in pre-enforced release times[2]. However it has not been much time since the fashion of short release cycles has come to the scene, affecting the Open source software market more than proprietary software market. The results first became visible when Canonical started its own version of Debian operating system with a 6 months release cycle instead of older average 4-5 year cycles of most of the operating systems. The results were very promising, Ubuntu soon emerged as the third most Used OS in desktops with the highest growth rate. Same has been continued by software such as Mozilla Firefox and Chromium. A study shows that Mozilla has not been able to keep up the Overall quality though the functionality has been improving noticeably [2]. On the other hand drastic downfall was observed when Banshee shortened their cycle; the company reverted back to their old release cycle. These varying results still leaves the question unanswered that how and with what external factors a shorter release cycle affects the quality and how exactly is cycle time related to readiness of software.

There is a very small literature available on understanding this scenario. Two noticeable papers: "Do Faster Releases Improve Software Quality? An Empirical Case Study of Mozilla Firefox" – Foutse Khom h, Tejinder Dhaliwal, Ying Zou, Bram Adams [2] and "Software release planning: an evolutionary and iterative approach" -D. Greer, G. Ruhe [1] may help us understand the current scenario. First one is a case study of Firefox and deals with quality estimation, second one tries to relate the incremental strategy to release decisions.

The problem of Software Release Planning dates back to early 80's. The Early solutions to this problem were fail proof as they insisted on limiting bugs to zero. One such fail-proof decision technique by Brettschneider R.[4] specifies a condition such that no test failures are permitted to be found in the a specified time limit before a release. Such a solution however no longer proves practical in today's business context. The aspect was Software Quality late 80's which narrowed down to Reliability. Various popular SRGMs( Software Reliability Growth Models) were proposed such as Jelinski-Morandal Model[5], NHPP Models, Exponential (Goel-Okumoto) Model[6], Modified Exponential Model etc..

These models were heavily used in software release time estimation in terms of saturation of a reliability factor. A sample work by W.Y. Yun and D.S. Bai used all these models for Release Estimation. [7].

In 90's software release planning became more business oriented and qualitative than ever. However the knowledge remained poorer. A few new approaches were used to model Release Planning Policies rather than estimating the time itself[8]. In next decade Release planning soon met field such as Data Mining & Soft Computing to solidify predictions. The most explanatory work in this era was "The Art and Science of Software Release Planning" [9], which tried to understand the problem with both qualitative and quantitative heads and human intuition.

Most of the works done so far used to estimate time using the data present in testing phase. However our aim was to estimate time during the requirement Analysis phase of incremental development. The only decision making data that might be present in this phase is feedback from previous phase as well as human intuition. We chose two popular works , EVOLVE[1] and Use Case Points[3] to which were directly in context with our problem and didn't require any testing data.

Since using a SGRM (Software Reliability Growth Model) was not possible in the Planning phase, so we have developed our own feedback mechanism and used it modify the EVOLVE approach.

It is very probable that in coming years more and more software will adhere to faster release cycles to cope up with the technology and competition. The trend is gaining popularity and needs to be thoroughly researched.

## 1.1 Evolve

EVOLVE is a proven evolutionary and iterative approach that optimally allocates requirements to increments and aims at continuous planning during incremental software development. We will use the EVOLVE to predict the requirements to be satisfied in the current iteration only. According to EVOLVE [1] $R^k$ is considered as a starting point to plan for increment k and represents a set requirements. $R^k$ is an n-tuple, and represents the set of n candidate requirements present on the beginning of a particular iteration.

E.g. $R^k = \{r1, r4, r6, r7, r8\}$, means that r3,r5 etc has already been implemented in previous iterations.

$S_p$ represents a stakeholder p and assigned a relative importance $\lambda_p \in (0,1)$ and normalized such that $\sum_{p=1\ldots q} \lambda_p = 1$. Where $\lambda$ is the calculated from the Eigen values [10] of the stakeholder comparison matrix. The method of averaging over normalized columns can be effectively used for Eigen values estimation.

Prio&Value are priority and business impact value determined by each $S_p$ for all requirements $\eta \in R^k$. They are represented as two dimensional matrices plotting stakeholders against requirements. Every intersection cell represents the priority/value of a particular requirement as perceived by the stakeholder. (We will modify these matrices in the proposed approach based on the feedback factor*)

$\omega^*$ is the assignment function which assigns each requirement to an increment $k, k+1$..
For all requirements $\eta \in R^k$ that are candidates for current iteration, Evolve defines the solution as an assignment $\omega^*(\eta) = s$, which satisfies all the following Equations( s being the increment). The following Equations are taken from EVOLVE and forms the problem statement. We will modify this approach in the next sub-section.

$$(1) \sum_{r(i)\in Inc(m)} effort(\eta, R^k) \leq size^m \, for \, m = k, k+1$$
(Effort Constraint)

This relation determines whether the size chosen (or maximum possible) is in accordance with the chosen requirements.

$$(2) \; \omega^*(\eta) \leq \omega^*(\eta_j) \, for \, all \, pairs \, (\eta, \eta_j) \in \varphi^k \, (Precedence \, Constraints)$$

A requirement having higher precedence must either be in same or previous increment than a requirement of lower precedence. $\varphi^k$ is the predefined set used to record the precedence constraints.

$$(3) \; \omega^*(\eta) = \omega^*(\eta_j) \, for \, all \, pairs \, (\eta, \eta_j) \in \rho^k$$
(Coupling Constraint)

Mutually dependent requirements must occur in same iteration. $\rho^k$ is the predefined set used to record the precedence constraints.

EVOLVE now defines two terms A (aggregated penalty) and B (aggregated benefit). A is an aggregated measure of wrong decisions made in choosing the requirements while B is an aggregated measure of correct decisions.

$$(4) \; A = \sum_{p=1\ldots q} \lambda_p [\sum_{r_i, r_j \in R(k)} penalty(\eta, \eta_j, S_p, R^k, \omega^*)]$$
$$=> \, min! \, with \, penalty(\eta, \eta_j, S_p, R^k, \omega^*) :=$$
$$(4.1) \, 0, \, if \, [prio(\eta, S_p, R^k) - prio(\eta_j, S_p, R^k)]$$
$$[\omega^*(\eta) - \omega^*(\eta_j)] > 0$$
$$(4.2) | \, prio(\eta, S_p, R^k) - prio(\eta_j, S_p, R^k)|$$
$$if \, \omega^*(\eta) = \omega^*(\eta_j)$$

$$(4.3) |\omega^*(r_i) - \omega^*(r_j)| \quad if$$
$$[prio(r_i, S_p, R^k) = prio(r_j, S_p, R^k)]$$
$$(4.4) [prio(r_i, S_p, R^k) - prio(r_j, S_p, R^k)]$$
$$[\omega^*(r_i) - \omega^*(r_j)] \ otherwise$$

Penalty hence calculated determines the aggregated sum of errors occurred by choosing a less priority requirement over higher priority requirement. This must be minimized.

Errors thus occurred are balanced by the Benefit factor.

$$(5) B = \sum_{p=1..q} \lambda_p [\sum_{r_i \in R(k)} benefit(r_i, S_p, \omega^*)] => max! \ with \ benefit(r_i, S_p, R^k, \omega^*) =$$
$$[\partial - value(r_i, S_p, R^k) + 1] * [\tau - \omega^*(r_j) + 1]$$
$$and \ \tau = max\{\omega^*(r_i): r_i \in R^k\}$$

Benefit hence calculated determines the aggregated sum of gains occurred by correctly choosing a higher value requirement over a lower one. This must be maximized.

The Penalty and benefit will then be added linearly to sum the total gains.

$$(6) C(\alpha) = (\alpha - 1)A + \alpha B => max \ \alpha \epsilon (0,1)$$

The overall linear sum of Benefit and Penalty is maximized.

$$(7) \ Determine \ K \ best \ solutions$$
$$from \ C(\alpha_1), C(\alpha_2), C(\alpha_3), with \ 1 \le K \le 10 \ and$$
$$0 \le \alpha_1 \le \alpha_2 \le \alpha_3 \le 1$$

Since we deal with all the combinations of requirements possible, a huge solution set is available and hence genetic algorithms can be successfully tried. A genetic algorithm is now applied to maximize objective function (6). Chromosomes satisfying 1-3 are the only valid solutions and hence are filtered and considered suitable for Genetic Algorithm. Ruhe suggests a crossover and mutation rate of 0.5 .The output is an assignment of all the requirements to increments.

In proposed model, the requirements assigned to current increment (release cycle) only will be of primary concern. The inputs, stakeholder-determined requirement-priority and requirement-value will be modified using a feedback mechanism discussed ahead. The Effort Constraints will be replaced with a time constraint. The next section describes the altered version of EVOLVE used in proposed approach

## 1.2 Altered EVOLVE

The EVOLVE model was primarily developed for requirements domain and doesn't deal in any way with time domain. Hence we needed to alter the model to make it suitable for time domain. We alter the EVOLVE method in two places to fit it in Time domain.

1.  He Effort Constraint is replaced by time constraint such that

$$\sum t_i \le T_{max}$$

Here $t_i$ represents the estimated time of a selected requirement. $T_{max}$ represents the Deadline Limit.

2.  The Prio and Value matrices are altered by multiplying the perceived values of all those requirements in $R^k$ which are being re-implemented (including the requirements generated as a consequence of previous requirement failures, e.g.: Major bugs) with inverse of the feedback factor i.e. $\frac{1}{FF}$, which will be introduced in further sections. In short, a feedback factor is an overall evaluation of model on 0-1 scale. The significance of feedback is discussed in the feedback factor section.

Time required for a pre-determined project can be best calculated in planning phase by Use Case Points. Time for individual requirements is then calculated using a weighted version of UCP discussed ahead

## 1.3 Use Case Points [3]

Use Case Points (UCP) is a widely-accepted use-case based software estimation approach. This technique was developed in 1993 by Gustav Karner primarily for object oriented systems and takes multiple technical and environmental considerations into account.

The equation is composed of four variables:

1. Technical Complexity Factor (TCF).
2. Environment Complexity Factor (ECF).
3. Unadjusted Use Case Points (UUCP).
4. Productivity Factor (PF).

Each variable is defined and computed separately, using perceived values and various constants. The complete equation is:   UCP = TCP * ECF * UUCP * PF

The UCP hence calculated is the estimated time for entire project considering that all the requirements will be implemented in a single increment. A solution for estimating time for each individual requirement is explained in the next section.

## 1.4 Weighted (Extended) Use case point's analysis

Consider r(1) to r(n) be all the candidate requirements that can be chosen for current release cycle. In a practical development scenario, we consider the requirements to be highly unique and specific and can be mapped to single use-cases. We consider a situation where all such requirements are needed to be implemented and apply the traditional UCP approach to determine a time T. If the number of requirements are n then,

Divide n requirements into three clusters, based on time needed (small, medium, big). Now assign proportional weights a, b, c respectively such that

- The value a/b, represents the approx ratio of time taken by small-size requirement to a medium-size requirement.
- The value b/c, represents the approx ratio of time taken by medium-size requirement to a big-size requirement.
- The value c/a, represents the approx ratio of time taken by big-size requirement to a small-size requirement.

Now let $$t = a + b + c$$

Let i, j, k be the respective number of requirement in small, medium and big size clusters.

$$n = i + j + k$$

The approximate time of a requirement is thus given by:

$$\frac{T*a}{(i*a+j*b+c*k)}, \text{for small size requirements}$$

$$\frac{T*b}{(i*a+j*b+c*k)}, \text{for medium size requirements}$$

$$\frac{T*c}{(i*a+j*b+c*k)}, \text{for big size requirements}$$

The Weights a, b, c can be conventionally assigned values 1, 2, 3 if a relative weight can't be estimated. Estimation can be further improved by using more than three clusters.

We now have set $T^k$ that holds respective times of requirements set $R^k$. We now calculate the feedback factor.

If we are in first increment we take the feedback factor

$$FF = 1$$

The value 1 signifies that feedback is either perfect or not yet available.
Else, the feedback factor is calculated with the pre-mentioned technique.
All those requirements in $T^k$ which are being re-implemented (including the requirements generated as a consequence of previous requirement failures, e.g.: Major bugs) are multiplied with inverse of the feedback factor i.e. $\frac{1}{FF}$.

We now introduce the feedback factor which is used to modify EVOLVE [1] inputs

## 1.5 FEEDBACK Factor

The reasons for not using the Software reliability growth models have already been explained. Instead a new approach is proposed to calculate the performance of our model and use this feedback as a mechanism to improve the future predictions and estimations of the model.
Let us define that (for immediate previous release)

- dT is a measure of difference in the estimated and actual time.
- FR represents the number of selected requirements which failed in some manner, i.e. not properly implemented, exceeded time by a huge amount , rejected by end users, faced a high count of bugs etc and needs to be re-implemented.
- User Perception (UP) is the rating of overall release by the end user or customer.

The method assumes that the variance or low feedback occurred because of one or more of following reasons:-

- Incorrect selection of requirements
- Incorrect priority or value Estimation by stakeholders
- Incorrect UCP time estimation

Hence we will now try to calculate a feedback factor (FF) which can be multiplied with the estimation values of requirements of previous release being re-implemented in current release. (It also applies to the newly generated requirements as a result of problems with previous release.)
A function Evl, which calculates the feedback factor is defined such that

$$FF = Evl(dT, FR, UP)$$

Evl is a linear function that sums up all the positive and negative feedbacks and gives a normalized output on 0-1 scale, 0 declaring a complete project failure and 1 declaring complete success.

It takes three inputs,

1. dT = 0, if actual time doesn't exceed the predicted time
       (T(actual)-T(estimated))/T(estimated), if actual exceeds the predicted time by a
       factor of two or less
   1, otherwise

2. FR = (total number of failed requirements)/ (total requirements implemented)
   It will range from 0 to 1. 0 being no failed requirements and 1 being the scenario where
   all requirements implemented failed.

3. UP is a customer rating [0-1], 0 being the minimum and 1 being maximum.
   Now we define

$$Evl\ (dT,\ FR,\ UP) = (UP - (0.5)(dT + FR))$$

The above formula gives 50% weightage to user perception & 50% weight to model accuracy (time & requirements) to calculate a normalized feedback factor. Importance percentages of user perception and model accuracy can be adjusted according to the nature of project and business environment.

FF (feedback factor) thus calculated will be used in proposed solution to Alter the UCP and EVOLVE inputs.

## 2. PROBLEM STATEMENT

Project X has just been started and is at verge of planning phase. The project has been declared feasible and all requirements are well defined and negotiated. The Project Manager has decided to deliver the requirements in an incremental fashion and needs to estimate the length of each release cycle. He asks all the stakeholders separately to prioritize and give a particular value to each requirement. Since all the stakeholders are not of same importance and caliber, he himself assigns relative importance to each one including himself. As the planning phase starts he now has the requirements mapped to discrete use cases. He now needs to estimate the project release cycle's using the limited available knowledge. This calls for the need of a decision support system to assist in required predictions.

## 3. PROPOSED SOLUTION

The solution is based on two assumptions. First, that choosing correct requirements helps in estimating the cycle time. Furthermore choosing correct requirements is directly influenced by performance of the model in previous increment, the ratio of failed requirements to total implemented and the user perception of each requirement.

The project manager now has a deadline to meet for current release; he decides a release cycle length. He needs a model to evaluate the decision as well as predict a best suited cycle time. A set of requirements is first determined and. Weighted Use case point's analysis is then performed to assign estimated time to each well-defined requirement. He now needs to decide which requirements to choose for the current release cycle. He uses the Altered EVOLVE model to achieve this.

Project Manager has the following inputs in hand

- Feedback factor from previous release (if any)
- Stakeholder priorities Matrix (Prio) for all requirements.

- Stakeholder Business Value Matrix (Value) for all requirements.
- Relative Stakeholder priorities
- Use Case Estimated time of each requirement
- Precedence Dependencies between requirements
- Coupling Dependencies between requirements
- A maximum deadline time (enforced by customer or higher management)

He can now proceed with the Altered EVOLVE method.

A random set of chromosomes is generated from candidate $R^k$ using the Subset-generation Algorithm. Hence each chromosome generated is a subset of power-set of $R^k$ (excluding Null Set). Hence for n requirements, the number of solutions generated is $2^n - 1$. This is a very large possible-solution set and contains many invalid solutions. We apply three constraints to filter out the invalid constraints.

- Time Constraint
- Precedence Constraint
- Coupling Constraint

Now with the valid solutions only in the possible-solution set, the Fitness function is calculated using the linear sum of Benefit and Penalty (6). Crossover and Mutation are performed at rates 0.5 each as suggested by Ruhe [1].

After sufficient GA iterations, a set of close solutions is obtained and a particular solution is manually chosen.

Time is then calculated as $\sum t_{selected}$.
Project then moves on to the next release cycle.

The Algorithmic steps of the proposed solution are briefly described as follows:

1. Determining a set of Requirements. A requirement can be a new feature, bugs or requirements not selected in previous releases. Each requirement must be map-able to unique use cases.
2. Calculate the Estimated time for each requirement using the Extended UCP method as explained
3. Calculate the feedback factor and multiply it with the selected requirements times.
4. Assign a time limit that must not be exceeded.
5. Input the Stakeholders data and their relative importance values. Use this matrix to calculate the Eigen Values.
6. Assign the stakeholder priorities and stakeholder values to each requirement for the current iteration.
7. Multiply the feedback factor to selected (repeated) stakeholder priorities and importance values as explained.
8. Determine the Coupling and Precedence constraints
9. Generate all possible Requirement sets using subset-generation algorithm.
10. Filter out the invalid chromosomes based on coupling, precedence and time constraints.
11. Assign a fitness value to each chromosome using objective function (6). Our aim is to maximize this function.
12. Randomly select 2 chromosomes from better half (having high fitness value) and perform crossover o generate new offspring.

13. Randomly select 2 chromosomes from better half (having high fitness value) and perform mutation o generate new offspring.
14. Add Offspring to population.
15. Go back to step 10, if more iterations needed (Population not yet converged)
16. Choose a best solution from new high fitness population.
17. Calculate the release cycle time.
18. If more iterations, determine failed requirements and resulting bugs. Go to Step-1.
19. Exit

Following flow diagram sums up the steps described in the preceding Algorithm in brief. The flow diagram represents the iterative nature of project as well as the proposed solution. A stopping condition has not been mentioned to represent an ideal incremental-condition such that project goes on. However the model stops as all the requirements are consumed and no major bugs are detected. The detailed implementation Algorithm is discussed in Appendix.

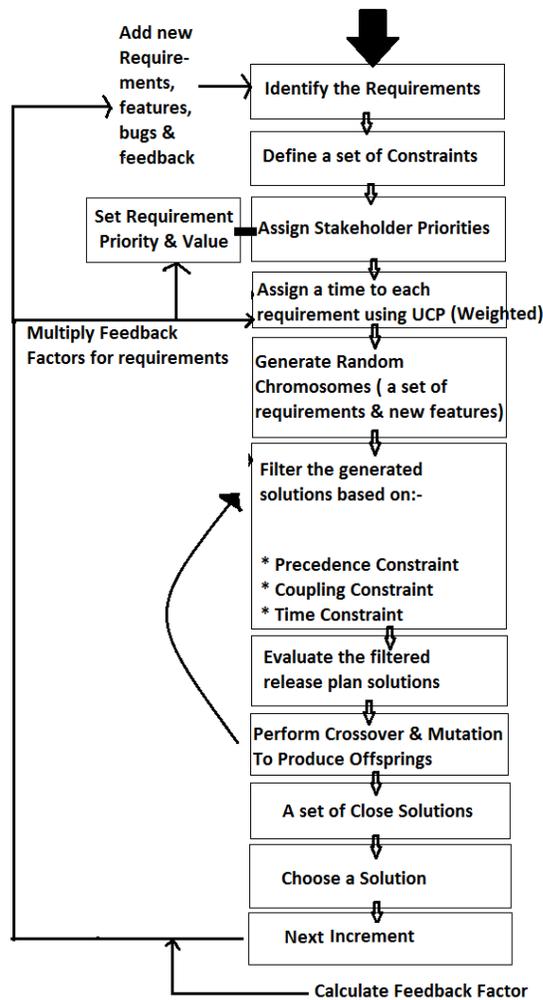

Figure 1 Proposed Solution - Flow of Steps

## 3. MODEL ANALYSIS CASE STUDIES

The solution is based on two assumptions. First, that choosing correct requirements helps in estimating the cycle time. Furthermore choosing correct requirements is directly influenced by performance of the model in previous increment, the ratio of failed requirements to total implemented and the user perception of each requirement.

*Description of Sample Project 1*

An online file storage service is to be implemented incrementally. The 7 Core Requirements to be coded are as follows:

1. Login Management
2. Session Management
3. Upload Module
4. Download Module
5. File Search Module
6. Sharing Management
7. Account Renewal

All these requirements pertain to the major use cases of the problem and hence Use Case Points analysis is applied.

All the values (factors) used below were carefully chosen on the basis of our own experience to suit the sample project as well as the college working environment.

Technical Complexity Factor (TCF) is estimated as follows:

Table 1 TCF Estimation

| Factors | Description | Weight | Perceived Complexity | Calculated Factor |
|---------|-------------|--------|----------------------|-------------------|
| T1 | Distributed System | 2 | 1 | 2 |
| T2 | Performance | 1 | 2 | 2 |
| T3 | End User Efficiency | 1 | 3 | 3 |
| T4 | Complex Internal Processing | 1 | 2 | 2 |
| T5 | Reusability | 1 | 2 | 2 |
| T6 | Easy to Install | 0.5 | 2 | 1 |
| T7 | Easy to Use | 0.5 | 3 | 1.5 |
| T8 | Portable | 2 | 1 | 2 |
| T9 | Easy to Change | 1 | 3 | 3 |
| T10 | Concurrent | 1 | 3 | 3 |
| T11 | Special security features | 1 | 4 | 4 |
| T12 | Provides direct access for third parties | 1 | 3 | 3 |

| T13 | Special user training facilities are required | 1 | 1 | 1 |
|-----|-----------------------------------------------|---|---|---|

Total Factor =29.5
TCP = 0.6 + (.01*Total Factor)  = 0.895
Environmental Complexity Factor (ECF) is estimated as follows:

Table 2 ECF Estimation

| Environmental Factor | Description | Weight | Perceived Impact | Calculated Factor |
|----------------------|-------------|--------|------------------|-------------------|
| E1 | Familiarity with UML | 1.5 | 1 | 1.5 |
| E2 | Application Experience | 0.5 | 1 | 0.5 |
| E3 | Object Oriented Experience | 1 | 1 | 1 |
| E4 | Lead analyst capability | 0.5 | 3 | 1.5 |
| E5 | Motivation | 1 | 3 | 3 |
| E6 | Stable Requirements | 2 | 3 | 6 |
| E7 | Part-time workers | -1 | 0 | 0 |
| E8 | Difficult Programming language | 2 | 1 | 2 |

Total Factors:15.5
ECF = 1.4 + (-0.03*Total Factor) = 0.935

Unadjusted Use Case Points (UUCP) is a sum of *Unadjusted Use Case Weight* (UUCW) and *Unadjusted Actor Weight* (UAW).  UUCW is estimated as follows:

Table 3 UUCW Estimation

| Use Case Type | Weight | Number of Use Cases | Result |
|---------------|--------|---------------------|--------|
| Simple | 5 | 3 | 15 |
| Average | 10 | 1 | 10 |
| Complex | 15 | 3 | 45 |

Total UUCW:70
UAW is estimated as follows:

Table 4 UAW Estimation

| Actor Type | Weight | Number of Actors | Result |
|---|---|---|---|
| Simple | 1 | 2 | 2 |
| Average | 2 | 2 | 4 |
| Complex | 3 | 1 | 3 |

Total UAW: 9
UUCP = UUCW + UAW = 79
PF (Productivity Factor) = 20 (Industry Average)
UCP = TCP * ECF * UUCP * PF = 1325 hours

Therefore total estimated time T: 1325 hours

Maximum time limit per release (say): 400 hours. This value depends upon the project but is always enforced by an authorizing stakeholder. Since we were supposed to complete the first phase of project in approximately 20 days with 5 stakeholders working around 4 hours per day, a value of 20*4*5 is taken as limit time.

The next step involves estimating approximate time for each requirement. Assuming 3 clusters of requirements with weights 1, 2,3 the estimation is calculated as follows:

Table 5 Estimating Requirement Time

| Cluster Type | Requirements | Number of Requirements | Weight | Time per Requirement (As predicted by Altered UCP) |
|---|---|---|---|---|
| Simple | 1,6,7 | 3 | 1 | 95 |
| Moderate | 2 | 1 | 2 | 189 |
| Complex | 3,4,5 | 3 | 3 | 283 |

Feedback Factor = 1, Since it's the first increment, hence no errors were occurred in previous increment (as it didn't exist), so Feedback factor becomes 1 .

Sample Stakeholder Assigned Values (on basis of their take on importance of each requirement on a 0-5 scale)

Table 6 Stakeholder Assigned Values

|  | R1 | R2 | R3 | R4 | R5 | R6 | R7 |
|---|---|---|---|---|---|---|---|
| S1 | 4 | 4 | 5 | 5 | 5 | 1 | 2 |
| S2 | 5 | 5 | 5 | 5 | 5 | 5 | 5 |
| S3 | 2 | 2 | 5 | 5 | 2 | 3 | 1 |
| S4 | 1 | 1 | 1 | 5 | 5 | 4 | 4 |
| S5 | 2 | 1 | 3 | 5 | 4 | 1 | 3 |

Sample Stakeholder Assigned Priorities (on basis of their take on priority of each requirement on a 1-7 scale)

Table 7 Stakeholder Assigned Priorities

|    | R1 | R2 | R3 | R4 | R5 | R6 | R7 |
|----|----|----|----|----|----|----|----|
| S1 | 1  | 2  | 3  | 4  | 5  | 6  | 7  |
| S2 | 1  | 3  | 2  | 5  | 4  | 6  | 7  |
| S3 | 1  | 3  | 4  | 5  | 6  | 2  | 7  |
| S4 | 1  | 4  | 5  | 6  | 2  | 3  | 7  |
| S5 | 1  | 4  | 5  | 6  | 2  | 3  | 7  |

Pair wise comparison of Stakeholders by Project Manager (In this case, Team Leader)

Table 8 Stakeholder Comparison

|    | S1   | S2   | S3   | S4 | S5 |
|----|------|------|------|----|----|
| S1 | 1    | 2    | 3    | 4  | 1  |
| S2 | 0.5  | 1    | 3    | 2  | 1  |
| S3 | 0.33 | 0.33 | 1    | 2  | 4  |
| S4 | 0.25 | 0.5  | 0.5  | 1  | 1  |
| S5 | 1    | 1    | 0.25 | 1  | 1  |

Requirement Precedence Dependency: {(R1,R2), (R1,R3),(R1,R6),(R1,R7)}

Requirement Coupling Dependency: {(3, 4)}

**Results**

The implementation software uses Genetic Algorithm Approach to pin down dominating solution sets. In most of the runs population converged at three highly fit solutions:
<R1>
<R1,R5>
<R1,R6>

We can now use our knowledge and logic to handpick one of them. We chose the <R1,R6> solution and calculated time by adding their individual estimated times.
Estimated release time for current release: 378 hours. This solution was in perfect coordination with our previous estimate as well as our actual project experience.

*Description of Sample Project 2(Industry Project)*

Sahara Bank, Libya (BNP Paribas Group) [11] needed to replace their legacy banking software in a quick incremental way. The Project was outsourced to TCS (Software Consultancy Organization) [12] and following modules were demanded from customer side.

1. Login Management
2. Scope Management

3. Admin Part
   3.1 Account Management
   3.2 Customer Management
   3.3 Employee Management

    3.4  ATM Management
    3.5  Brach or Bank Management
    3.6  Region Management

4. Customer Part
    4.1  ATM Banking
    4.2  NET Banking
    4.3  Core Banking
    4.4  Phone Banking

The stipulated time for project was three weeks and a team of 27 members worked on the project. The project was delivered successfully in three quick increments within the stipulated time. First increment was released for beta testing on 9th day of project, second on 15th and final increment on 20th day. The feedback was highly positive for all three incremental releases.

As per the data provided by Tech Lead of Project, the following major use cases were determined and later implemented.

1.  Login Management
2.  Scope Management
3.  Customer and employee interface interaction for atm banking, core banking, net banking,   and phone banking
4.  Create, view, view all, update, delete ,deactivate and activate region ,branch ,atm, customer, employee, account etc.
5.  Fund transfer from region to branch, branch to sub-branches and atms' in morning and evening accounting into threshold balance
6. Interest calculation
7. Cheque-book request & processing
8. Fund transfer from one account to another, Bill Payment
9. Foreign Currency exchange
10. Account, Balance and transaction limits
11  Validations -both back end and front end

All the values (factors) given below reflects the nature of requirements by Sahara Bank and are assigned by Tech Lead on basis of his perception of project. (Note: No UCP Analysis was carried out during the project and the following perceived complexity factors have been determined by Project team to facilitate the analysis of our research work)

Technical Complexity Factor (TCF) is estimated as follows:

Table 9 TCF Estimation

| Factors | Description | Weight | Perceived Complexity | Calculated Factor |
|---------|-------------|--------|---------------------|-------------------|
| T1 | Distributed System | 2 | 0 | 0 |
| T2 | Performance | 1 | 4 | 4 |
| T3 | End User Efficiency | 1 | 4 | 4 |
| T4 | Complex Internal Processing | 1 | 1 | 2 |

| T5 | Reusability | 1 | 2 | 2 |
|----|-------------|---|---|---|
| T6 | Easy to Install | 0.5 | 1 | 0.5 |
| T7 | Easy to Use | 0.5 | 4 | 2 |
| T8 | Portable | 2 | 1 | 2 |
| T9 | Easy to Change | 1 | 3 | 3 |
| T10 | Concurrent | 1 | 3 | 3 |
| T11 | Special security features | 1 | 5 | 5 |
| T12 | Provides direct access for third parties | 1 | 5 | 5 |
| T13 | Special user training facilities are required | 1 | 0 | 0 |

Total Factor =32.5

TCP = 0.6 + (.01*Total Factor)   = 0.925

Environmental Complexity Factor (ECF) is estimated as follows:

Table 10 ECF Estimation

| Environmental Factor | Description | Weight | Perceived Impact | Calculated Factor |
|----------------------|-------------|--------|------------------|-------------------|
| E1 | Familiarity with UML | 1.5 | 2 | 3 |
| E2 | Application Experience | 0.5 | 2 | 1 |
| E3 | Object Oriented Experience | 1 | 4 | 4 |
| E4 | Lead analyst capability | 0.5 | 4 | 2 |
| E5 | Motivation | 1 | 4 | 4 |
| E6 | Stable Requirements | 2 | 4 | 8 |
| E7 | Part-time workers | -1 | 0 | 0 |
| E8 | Difficult Programming language | 2 | 0 | 0 |

Total Factors: 22
ECF = 1.4 + (-0.03*Total Factor) = 0.740
UUCW is estimated as follows:

Table 11 UUCW Estimation

| Use Case Type | Weight | Number of Use Cases | Result |
|---|---|---|---|
| Simple | 5 | 4 | 20 |
| Average | 10 | 4 | 40 |
| Complex | 15 | 3 | 45 |

Total UUCW: 105

UAW is estimated as follows:

Table 12 UAW Estimation

| Actor Type | Weight | Number of Actors | Result |
|---|---|---|---|
| Simple | 1 | 3 | 3 |
| Average | 2 | 4 | 8 |
| Complex | 3 | 1 | 3 |

Total UAW: 14

UUCP = UUCW + UAW = 119

PF (Productivity Factor) = 24 (Estimated TCS Average)

UCP = TCP * ECF * UUCP * PF = 1955 hours

Therefore total estimated time T: 1955 hours

Maximum time limit per release: 1300 hours. This value is representative of the time constraints enforced by Sahara Bank on TCS team for first review of Project.

The next step involves estimating approximate time for each requirement. Assuming 3 clusters of requirements with weights 1, 2, 3 the estimation is calculated as follows:

Table 13 Estimating Requirement wise time

| Cluster Type | Requirements | Number of Requirements | Weight | Time per Requirement (As predicted by Altered UCP) |
|---|---|---|---|---|
| Simple | 1,7,9,10 | 4 | 1 | 70 |
| Moderate | 4,5,6,8 | 4 | 2 | 140 |
| Complex | 2,3,11 | 3 | 3 | 211 |

Feedback Factor = 1, since it's the first increment, hence no errors were occurred in previous increment (as it didn't exist), so Feedback factor becomes 1.

Eight Stakeholders including the domain expert from customer side were chosen such that they represent the entire project team of 27 members.

Sample Stakeholder Assigned Values (on basis of their take on importance of each requirement on a 0-5 scale and are representative of various streams of thoughts of the stakeholders)

Table 14 Stakeholder Assigned Values

|    | R1 | R2 | R3 | R4 | R5 | R6 | R7 | R8 | R9 | R10 | R11 |
|----|----|----|----|----|----|----|----|----|----|-----|-----|
| S1 | 4  | 3  | 5  | 4  | 4  | 2  | 2  | 3  | 2  | 2   | 4   |
| S2 | 3  | 3  | 4  | 4  | 4  | 3  | 3  | 3  | 2  | 2   | 3   |
| S3 | 3  | 2  | 5  | 5  | 2  | 3  | 3  | 3  | 3  | 3   | 2   |
| S4 | 3  | 3  | 3  | 5  | 5  | 3  | 3  | 4  | 3  | 2   | 2   |
| S5 | 3  | 2  | 5  | 5  | 2  | 3  | 3  | 3  | 3  | 3   | 2   |
| S6 | 4  | 3  | 5  | 4  | 4  | 2  | 2  | 3  | 2  | 2   | 4   |
| S7 | 3  | 3  | 4  | 4  | 4  | 3  | 3  | 3  | 2  | 2   | 3   |
| S8 | 4  | 3  | 5  | 4  | 4  | 2  | 2  | 3  | 2  | 2   | 4   |

Sample Stakeholder Assigned Priorities (on basis of their take on priority of each requirement on a 1-11 scale and are representative of various streams of thoughts of the stakeholders)

Table 15 Stakeholder Assigned Priorities

|    | R1 | R2 | R3 | R4 | R5 | R6 | R7 | R8 | R9 | R10 | R11 |
|----|----|----|----|----|----|----|----|----|----|-----|-----|
| S1 | 1  | 11 | 2  | 4  | 9  | 8  | 7  | 6  | 10 | 5   | 3   |
| S2 | 1  | 11 | 2  | 5  | 4  | 6  | 7  | 9  | 10 | 8   | 3   |
| S3 | 1  | 11 | 2  | 3  | 9  | 8  | 7  | 6  | 10 | 5   | 4   |
| S4 | 1  | 11 | 2  | 4  | 9  | 8  | 7  | 6  | 10 | 5   | 3   |
| S5 | 1  | 11 | 2  | 5  | 4  | 6  | 7  | 9  | 10 | 8   | 3   |
| S6 | 1  | 11 | 2  | 4  | 9  | 8  | 7  | 6  | 10 | 5   | 3   |
| S7 | 1  | 11 | 2  | 5  | 4  | 6  | 7  | 9  | 10 | 8   | 3   |
| S8 | 1  | 11 | 2  | 4  | 9  | 8  | 7  | 6  | 10 | 5   | 3   |

Pair wise comparison Of Stakeholders by Project Manager (In this case, Team Leader has determined the values)

Table 16 Stakeholder Comparison

|    | S1   | S2   | S3  | S4 | S5  | S6 | S7 | S8 |
|----|------|------|-----|----|-----|----|----|----|
| S1 | 1    | 1    | 3   | 3  | 4   | 4  | 3  | 3  |
| S2 | 1    | 1    | 3   | 2  | 3   | 3  | 2  | 2  |
| S3 | 0.33 | 0.33 | 1   | 2  | 2   | 2  | 2  | 2  |
| S4 | 0.33 | 0.5  | 0.5 | 1  | 1   | 1  | 1  | 1  |
| S5 | 0.25 | 0.33 | 0.5 | 1  | 1   | 1  | 2  | 2  |
| S6 | 0.25 | 0.33 | 0.5 | 1  | 1   | 1  | 1  | 1  |
| S7 | 0.33 | 0.5  | 0.5 | 1  | 0.5 | 1  | 1  | 1  |
| S8 | 0.33 | 0.5  | 0.5 | 1  | 0.5 | 1  | 1  | 1  |

Requirement Precedence Dependency:
{(1, 3), (1, 11)}
Requirement Coupling Dependency:
{(3, 11)}

## Results

We tested the project various time on our implementation and found that all requirements are consumed in 3 to 4 iterations depending upon the feedback factor and requirements chosen.
From second iteration we considered a feedback of 0.8 to 0.9 which was representative of the highly positive feedback from Sahara Team in each review.
Following Solutions were converged in first iteration.
<R1,R3,R4,R11>
<R1, R11,R3>
<R1,R3,R11>
Choosing one of these solutions determined the number of further iterations.

The results were in accordance with TCS original scenario, where 3 iterations were done such that following requirements were implemented.

Iteration-1: R1, R3, R11
Iteration-2: R4, R10, R8, R6, R7
Iteration-3: R2, R5, R9

We also found that a positive a feedback tends towards reducing the number of iteration, a detailed analysis of this result has been done in next section.

The Results we received in various runs were highly coherent with the actual TCS Project experience. Fig-2 shows a comparison of various runs of proposed solution with the actual results. Various runs assumed different values of feedback factor ranging from 0.75 to 0.9 (depicting a highly positive feedback by client) and slight variations were deliberately done in choosing the solution set to check the robustness.

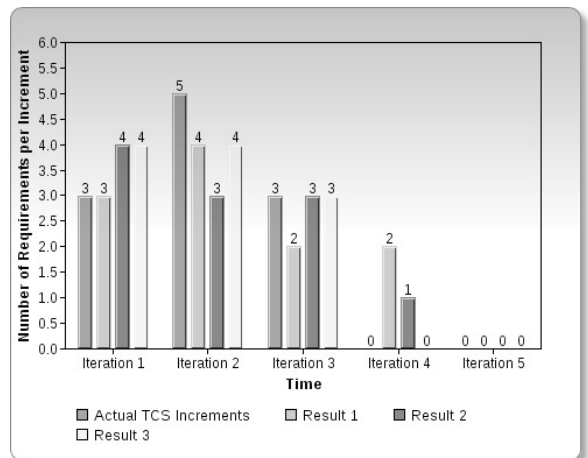

Figure 2 Comparison of Results

In above comparison, the first bar of each iteration depicts the actual TCS results followed by our results in various runs. It was interesting to see that no solutions suggested a fifth iteration. Result-1 assumed a feedback factor of 0.9 and was most coherent with actual results. Result-2 and Result-3 were determined at a lower feedback and hence led to more iterations.  Such a coherency with TCS Project confirmed the accuracy and robustness of the proposed solution.

# 4. MODEL COMPUTATIONAL ANALYSIS

The implementation method was tested on a i3, 2$^{nd}$ generation machine and it was found that the proposed solution becomes more and more memory-hungry as the number of requirements increase beyond a saturation limit. Hence a parallel & distributed implementation of the solution is advised. Fig-3 shows the tradeoff between number of requirements and time complexity. Fig-3 was extrapolated and interpolated to suit a complete requirement range. We detected an exponential growth.

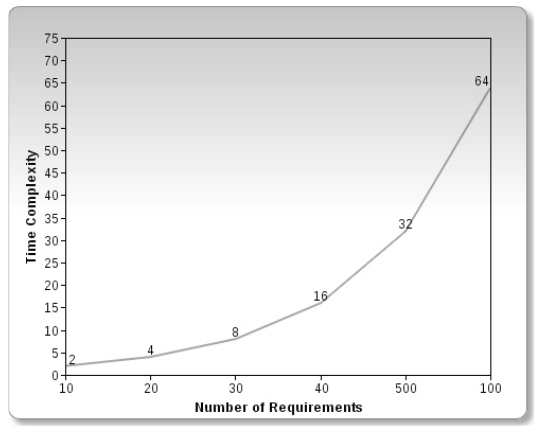

Figure 3 Requirements vs. Time

Coming to feedback factor, very positive results were observed. As the number of iterations (increments) increases, the feedback factor decreases to a certain limit. This confirms that project might be going in right direction, however as the number of increments increase beyond a certain limit (which signifies that more and more bugs & failures are being encountered), the feedback keeps on decreasing towards zero, confirming a failed project. Fig-4 was extrapolated and interpolated to suit a complete requirement range based on 12 observations on sample projects.

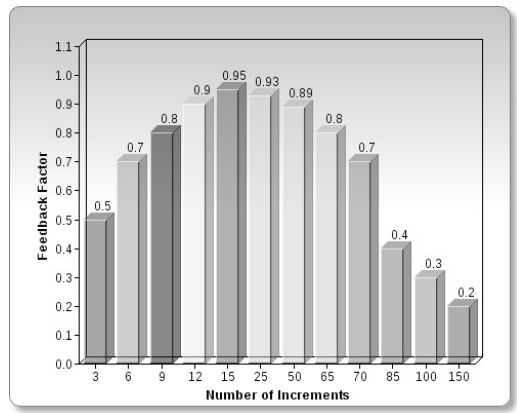

Figure 4 Feedback Factor vs. Number of Increments

## 5. CONCLUSIONS

In both the projects discussed above, time estimated matches the estimated time as well as close to time taken in implementing the real project. Hence the model seems to suit both small (college projects) as well as the large industry projects. However, the idea proposed has a lot of scope for improvement, the factors considered in Use Case points can be studied further and necessary alterations can be made to suit certain project types in general. The solution for now uses Subset generation algorithm which demands very high computation as the project requirements and bugs increases rapidly. From analysis, we can see that it might be difficult to handle projects that need a complete reengineering. We must consider the need and solution for implementing the approach on a parallel (or distributed) system to account for computational problems.

## APPENDIX

**Algorithm**

BEGIN

  FF: = getFEEDBACK();  // calculates feedback
  UCP: = CalcUCP();  // calculates UCP
  Assign $T^R$ ();    // Extends UCP
  Get-Matrix(Stakes);  //Gets relative importance
  Get-Eigen (Stakes);  //Calculates Eigen Values
  Get-Matrix (Prio);  // Gets Stakeholder priorities
  Get-Matrix (Value);  // Gets Stakeholder values
  Feed-Matrix (Prio);  // Multiplies Feedback
  Feed-Matrix (Value);
  Get-Precedence(Prec); // Gets Requirements Precedence
  Get-Coupling(Coup); // Gets Requirements Coupling
  Solution list [] [] = get-Subsets ( $R^R$ )
  //Generates Subsets
  Loop(n)

  For Each Element in Solution List[][]
   If Check_Prec(Element,Prec) = False || Check_Coup(Element, Coup) = FALSE  then
    Delete ELEMENT;
   End If
  Next Element
 For Each Element in Solution List[][]
   Calc_Fitness(Element)
  Next Element

  Sort_List ( Solution List[][])

 For Each E1,E2 in Solution List[][]
   E1=Select_Element(Solution_List);
   E2=Select_Element(Solution_List);
   New_Solution_String = Crossover(E1,E2)
   New_Solution_String = Mutation (New_Solution_String)
   Solution_List= New_Solution_String
 Next E1,E2

  If  (Converged Solution < X) , EXIT

End Loop

Choose Element;
Determine_Time(Element, $T^k_i$)

END

## REFERENCES

[1]   Greer D, Ruhe G, Software release planning: an evolutionary and iterative approach,    Information and Software Technology journal, Volume 46, Issue 1, pp. 243-253, Elsevier 2004

[2]   Khom F, Dhaliwal T, Zou Y, Adams B, Do Faster Releases Improve Software Quality? An Empirical Case Study of Mozilla Firefox, 9th IEEE Working Conference on Mining Software Repositories (MSR), Zurich, pp. 179 - 188 , 2012

[3]   Karner G, Use Case Points - Resource Estimation for Objectory Projects , Objective Systems SF AB (copyright owned by Rational Software), 1993

[4]   Brettschneider R, Is your Software Ready for Release, IEEE Software, Volume 6, Issue 4, 1989

[5]   Jelinski, Z. Moranda, P.B., Jelinski-Morandal Model, Statistical Computer Performance Evaluation,. Academic Press, New York, pp. 465-484, 1972

[6]   Okumoto K, Goel A L, Optimum releases time for software system based on reliability and cost criteria, Journal of Systems and Software, Elsevier,  Volume 1, pp. 315-318, 1984

[7]   W.Y. Yun and D.S., Bai, Optimum Software Release Policy with Random Life Cycle, IEEE, June 1990

[8]   Hoek A, Hall R S, Heimbigner D, Wolf A L, Software Release Management, ACM SIGSOFT Software Engineering Notes, Volume 22, Issue 6, pp. 159-175, 1997

[9]   Ruhe G, Saliu M.O , The Art and Science of Software Release Planning, IEEE Software, Volume  22, Issue 6, pp.47-53,  2005

[10]  Wang Q, Lai X, Requirements   Management for the Incremental Development Model, Proceedings of Second Asia-Pacific Conference on Quality Software, Hong Kong  pp. 295-301, 2001

[11]  Sahara Bank, Libya (BNP Paribas Group), www.saharabank.com.ly, www.bnpparibas.com

[12]  Tata Consultancy Services, www.tcs.com